\documentclass[aps,pre,twocolumn,a4paper,byrevtex,superscriptaddress,showpacs,showkeys,bibnotes,longbibliography]{revtex4-1}
\usepackage{microtype}
\usepackage{graphicx}
\usepackage{amsmath}
\usepackage{amssymb}
\usepackage{color}
\definecolor{myblue}{rgb}{0.153,0.322,0.706}
\usepackage[colorlinks,linkcolor=myblue,urlcolor=myblue,citecolor=myblue]{hyperref}

\newcommand{\be}{\begin{equation}}
\newcommand{\ee}{\end{equation}}
\newcommand{\ra}{\rightarrow}

\newcommand{\tG}{\tilde{G}}
\newcommand{\tU}{\tilde{U}}
\newcommand{\tV}{\tilde{V}}

\begin{document}
\title{Phase transitions in large deviations of reset processes}

\author{Rosemary J. Harris}
\address{School of Mathematical Sciences, Queen Mary University of London, London E1 4NS, UK}
\email{rosemary.harris@qmul.ac.uk}

\author{Hugo Touchette}
\address{National Institute for Theoretical Physics (NITheP), Stellenbosch 7600, South Africa}
\address{Institute of Theoretical Physics, Department of Physics, University of Stellenbosch, Stellenbosch 7600, South Africa}
\email{htouchette@sun.ac.za, htouchet@alum.mit.edu}

\begin{abstract}
We study the large deviations of additive quantities, such as energy or current, in stochastic processes with intermittent reset. Via a mapping from a discrete-time reset process to the Poland-Scheraga model for DNA denaturation, we derive conditions for observing first-order or continuous dynamical phase transitions in the fluctuations of such quantities and confirm these conditions on simple random walk examples. These results apply to reset Markov processes, but also show more generally that subleading terms in generating functions can lead to non-analyticities in large deviation functions of ``compound processes'' or ``random evolutions'' switching stochastically between two or more subprocesses.
\end{abstract}

\date{\today}


\maketitle

\section{Introduction}

There has been a renewed interest recently in stochastic processes involving random resets to a fixed state, representing, for example, the reduction of a population after a catastrophe \cite{pakes1978,brockwell1985,kyriakidis1994,pakes1997,dharmaraja2015}, the random attachment of a molecular motor on a biological filament \cite{meylahn2015}, the clearing of a queue or buffer \cite{crescenzo2003}, or a random search reinitialized to its starting position \cite{evans2011,evans2011b,evans2013,eule2016,kusmierz2014,janson2012,benichou2007}. The focus in random search applications is on the mean first-passage time, which can be optimized under reset \cite{evans2011b,evans2013,eule2016}, but other important statistical quantities have also come to be studied, including time-dependent distributions \cite{brockwell1982,kumar2000,crescenzo2012,majumdar2015b}, moments \cite{dharmaraja2015}, and large deviation functions \cite{meylahn2015b}.

Our goal in this letter is to continue the study of large deviations for Markov processes with reset \cite{meylahn2015b}. We consider as a general framework a Markov process $X_i$ evolving in discrete time, possibly with weakly time-dependent transition probabilities, and an observable $J_n$ of that process integrated over $n$ time steps. To be concrete, we will refer to $J_n$ as a ``current'' (e.g., of an interacting particle system) but other quantities can also be considered, such as the time a random walker spends in a given region or the work that a molecular motor expends over time as it moves on a filament before its position is reset. What is important in each case is that $J_n$ is not incremented in time, but simply retains its value whenever the process $X_i$ is reset (e.g., by returning particles to a particular configuration or restarting an internal clock).  

If the time between reset events is finite, then one anticipates that finite-time contributions to the generating function of $J_n$ for the subprocess \emph{without} reset play a role in determining the asymptotic behaviour of the generating function for the full process \emph{with} reset. Indeed, one expects that a specific current fluctuation will be optimally realised by a particular reset frequency. An interesting question is whether this optimal frequency depends smoothly on the value of the current fluctuation or whether there is a phase transition to a regime where current fluctuations are optimally realised by trajectories involving no reset at all. 

To answer this question, we show that the generating function of $J_n$ can be mapped to the partition function of the Poland-Scheraga (PS) model for DNA denaturation \cite{poland1966b,poland1966,kafri2000,richard2004} and use this mapping to derive criteria for first-order (discontinuous) and second-order (continuous) phase transitions in the large deviation functions describing the fluctuations of $J_n$. The results can be applied in principle to any additive observables of Markov processes; in this brief study, we focus on illustrating the approach for simple random walk models and then discuss other potential applications in the concluding section.

\section{Framework and results}

We are concerned with characterizing the fluctuations of $J_n$ for large integration times. For many systems and observables of interest, especially if the transition probabilities are time homogeneous or only weakly time inhomogeneous, the distribution of $J_n$ without reset has the large deviation form,
\be
P(J_n/n=j) \approx e^{-nI_0(j)}
\label{e:ldp1}
\ee
in the limit of large $n$ \cite{dembo1998,touchette2009,harris2013}. In this case, the distribution is fully characterized, up to subleading corrections in $n$, by the rate function
\be
I_0(j)=\lim_{n \ra \infty} -\frac{1}{n}\ln P(J_n/n=j),
\label{e:RF0}
\ee
written with the subscript $0$ to indicate that it is obtained for the process without reset.

Instead of considering the distribution and its associated rate function, we can also consider the generating function defined as
\be
G_0(k,n)=\langle e^{kJ_n} \rangle_0, 
\label{e:G0}
\ee
where the angular brackets denote an average over stochastic trajectories, started from some given initial distribution, and the subscript $0$ refers again to the original process without reset. The exponential scaling (\ref{e:ldp1}) implies that the generating function also scales exponentially
\be
G_0(k,n)\approx e^{n\lambda_0(k)},
\ee
with an exponent
\be
\lambda_0(k)=\lim_{n \ra \infty} \frac{1}{n} \ln G_0(k,n). 
\label{e:SCGF0}
\ee
referred to as the scaled cumulant generating function (SCGF). Moreover, it is known from the G\"artner-Ellis Theorem \cite{dembo1998} that, if $\lambda_0(k)$ is differentiable, then $I_0(j)$ can be obtained as the Legendre-Fenchel (LF) transform of the SCGF:
\be
I_0(j)=\max_k\{kj-\lambda_0(k)\}.
\ee
For non-differentiable $\lambda_0(k)$, the transform above gives only the convex hull of $I_0(j)$ and further arguments are needed to determine its true shape (see \cite{touchette2009}, Sec. 4.4).

Our aim now is to determine how the above large deviation functions, in particular the SCGF, change under the addition of reset.  To this end, we consider a reset version of the process $X_i$ which, at each time step, has a probability $r$ to be reset (with no current flowing) and which evolves otherwise according to its ``natural'' dynamics (with the current correspondingly incremented). We emphasize that the reset event does not reset the current -- it only resets $X_i$ by returning it to a given initial position (or distribution) or by restarting an internal clock variable at zero.  Finite-time corrections to~\eqref{e:G0} might then determine the form of the generating function for the whole ``compound'' process with reset which we denote as
\be
G_r(k,n)=\langle e^{kJ_n}\rangle_r.
\ee

\begin{figure}[t]
\centering
\resizebox{3.4in}{!}{\includegraphics{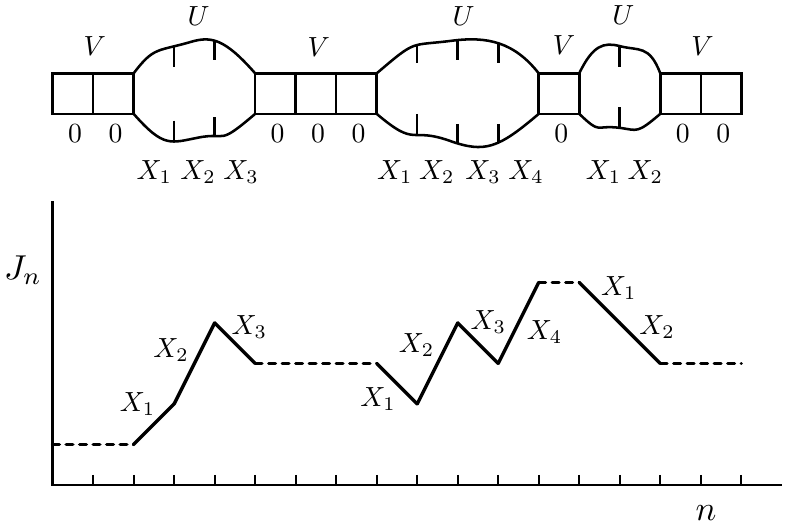}}
\caption{Relation between the current $J_n$ under reset and the Poland-Scheraga model of DNA denaturation (monomer pairs schematically indicated between vertical lines). Periods with reset, where $J_n$ is not incremented in time, correspond to bonds in the DNA model, whereas periods without reset, where $J_n$ evolves in time, correspond to denatrured DNA loops.}
\label{figPS}
\end{figure}

Conceptually, this problem can be considered as a temporal analogue of the Poland-Scheraga (PS) model for DNA denaturation in which a double-stranded chain is formed of pairs of bound and unbound monomers \cite{poland1966b,poland1966,kafri2000,richard2004}. Loops of denatured DNA (i.e., consecutive unbound monomers) can be mapped in our framework to temporal periods without reset whereas bound monomers correspond to reset events, as shown in Fig.~\ref{figPS}. In the PS model one calculates the partition function and looks for phase transitions (as a function of temperature) with the fraction of bound monomers as order parameter.  We here perform an analogous analysis for the current generating function, looking for phase transitions as a function of the conjugate parameter $k$.

Following the PS approach, we write the current generating function for a ``loop'' of $n$ consecutive steps without reset as
\be
U(k,n) = (1-r)^n G_0(k,n)
\ee
and that for a period of $n$ consecutive reset steps as
\be
V(k,n)=r^n e^0 = r^n.
\ee
The latter equation reflects the fact that the current is not incremented during a reset event; see again Fig.~\ref{figPS}.  One could also use the same formalism with more general generating functions to analyse switching between two or more different stochastic ``subprocesses'' that each increment the current but with different probability distributions (similar to the sequences of several types discussed in~\cite{lifson1964}).

As in the PS model, it is convenient to consider discrete-Laplace-transformed  (``$z$-transformed'') generating functions \cite{lifson1964}:
\begin{align}
\tG_r(k,z)&=\sum_{n=1}^\infty G_r(k,n) z^{-n}, \label{e:Grt} \\
\tU(k,z)&=\sum_{n=1}^\infty U(k,n) z^{-n}, \\
\tV(k,z)&=\sum_{n=1}^\infty V(k,n) z^{-n} = \frac{r}{z-r}.
\end{align}
This amounts to working within a grand-canonical ensemble in time (with fugacity $z^{-1}$) so the total number of steps fluctuates rather than being constrained to a constant.  

The $z$-transformed generating function $\tG_r$ for the full reset process can be written down explicitly by observing that its trajectories consist of alternating segments of consecutive no-reset steps and consecutive reset steps, leading to a geometric sum of $\tU\tV$ terms that yields
\be
\tG_r(k,z) = \frac{ \tU(k,z) + \tV(k,z) + 2\tU(k,z)\tV(k,z)}{1-\tU(k,z)\tV(k,z) }. 
\label{e:G}
\ee
This has essentially the same form as the grand-canonical (in space) partition function for the PS model with the precise numerator depending on our exclusion of zero-length trajectories (chains) and choice of free boundary conditions.

Note that, for the special case where the current increments for each step (in the original process without reset) are independent and identically distributed, there are no finite-time corrections in the generating function and one has the exact relation
\be
G_0(k,n)=e^{n\lambda_0(k)},
\ee
which straightforwardly leads from (\ref{e:G}) to
\be
\tG_r(k,z)=\sum_{n=1}^\infty\, [r+(1-r)e^{\lambda_0(k)}]^n z^{-n}.
\ee
Hence, by inspection,
\be
G_r(k,n)= [r+(1-r)e^{\lambda_0(k)}]^n,
\ee
and the SCGF is
\be
\lambda_r(k)=\ln[r+(1-r)e^{\lambda_0(k)}].
\ee
This reflects the obvious fact that, in this case, the generating function for a single step in the compound reset process is a weighted sum of the generating functions for a single step in the two subprocesses.   An example here is to take as original process a random walker which steps one lattice unit right with probability $p$ and one unit left with probability $(1-p)$ and so has current generating function
\be
G_0(k,n)=pe^{k}+(1-p)e^{-k}.
\ee  
The reset might force the random walker back to a particular point on the lattice but has no direct effect on the current counting, so the compound process must have the same generating function as a ``lazy'' random walk with probability $r$ to not move.

In more general cases, we can determine the SCGF of the reset process by locating, as in the PS model, the largest real value of $z$ at which $\tG_r(k,z)$ of (\ref{e:G}) diverges. In the absence of a phase transition, we thus look for the largest real solution of
\be
\tU(k,z) \tV(k,z) = 1,
\ee 
denoted by $z^*(k)$.  With the explicit form for $\tV(k,z)$, we have
\be
\tU(k,z^*) = \frac{z^*(k)}{r} - 1 
\label{e:zstar}
\ee
which is identical to the corresponding condition in the PS model (see, e.g.,~(5) in~\cite{poland1966}) when the reset probability $r$ is identified with the statistical weight of a bound pair \footnote{Since, by construction, $\tU(k,z)>0$ for positive $z$ we must have $z^*>r$ so $z^*$ is inside the region of convergence of $\tV(k,z)$.}. Since $z^*$ determines the leading (exponential) behaviour of $G_r(k,n)$ in $n$, we then obtain
\be
\lambda_r(k)=\ln z^*(k).
\label{eqscgfres1}
\ee
A phase transition occurs in this context when for some value of $k$ the function $z^*(k)$ reaches the convergence boundary point $z_c(k)$ of $\tU(k,z)$. Denoting this transition point as $k_c$, we must therefore have \footnote{The form of this equation also ensures $z_c(k) > r$.} 
\be
\tU(k_c,z_c) = \frac{z_c(k_c)}{r} - 1 
\label{e:kc}.
\ee

Analysis of the PS model reveals that the existence and nature of a phase transition is determined by the behaviour of $\tU(k,z)$ in the neighbourhood of $z_c$, which itself depends on the leading (exponential) and subleading (power-law) terms in the long-time limit of $U(k,n)$ \footnote{This holds provided $\lambda_0(k)$ and $c(k)$ are smooth functions of $k$; a phase transition could also arise from either or both functions being non-analytic in $k$.}. To be specific, let us assume the general scaling
\be
U(k,n) \sim \frac{(1-r)^n e^{n\lambda_0(k)}}{n^{c(k)}}, 
\label{e:Unlongtime}
\ee
so that $z_c(k)=(1-r) e^{\lambda_0(k)}$. If there is a phase transition at a finite value $k_c$, then $\tU(k_c,z_c)$ must converge, meaning $c(k_c)>1$. From there, the nature of the phase transition is determined as in the PS model by the value of the exponent $c$ \cite{poland1966b,poland1966,kafri2000,richard2004}:

i) For $1<c(k_c)\leq 2$, the derivative $\partial \tU(k,z)/\partial z$ diverges at $k_c$, leading to a continuous dynamical phase transition;

ii) For $c(k_c)>2$, the derivative converges at $k_c$, leading to a first-order or discontinuous dynamical phase transition with a ``cusp'' (discontinuous slope in $k$) in $\lambda_r(k)$.

A slight subtlety here is that, in the original PS model, the exponent $c$ and the value of $z_c$ are constants but the weight of a bound pair depends on temperature.  In contrast, in our framework the reset probability is constant but both $c$ and $z_c$ depend on the parameter $k$ conjugate to the current $J_n$. This does not change the PS criteria for first-order and continuous phase transitions, since these are based only on the convergence of $\tU$ and its derivative at specific parameter values  \footnote{The convergence of equivalent sums determines condensation transitions in Bose-Einstein gases, as pointed out in~\cite{poland1966}, and in zero-range processes~\cite{evans2005b}.}.

Physically, a phase transition is here between a regime where the current fluctuations are optimally realised by reset events at a finite fraction of time steps, leading to \eqref{eqscgfres1}, and a regime where the current fluctuations are optimally realised by trajectories with no reset events, so that
\be
\lambda_r(k)=\ln z_c(k) = \lambda_0(k)+\ln(1-r),
\label{eqscgfnr1}
\ee 
the $1-r$ factor obviously accounting for the probability of seeing no reset. If the phase transition is first-order, the average length of a segment without reset is finite at the transition point.  

It is worth emphasizing that the value of $c(k_c)$ in a particular reset scenario will in general depend on $r$: the nature of any phase transition can depend sensitively on the reset probability, in contrast to the PS model where the nature of the transition is \emph{a priori} independent of the binding weight.  We will see examples of this dependence in the following section.

\section{Examples}

Observables that are incremented in time with independent and identically distributed random variables have purely exponential generating functions ($c=0$), as seen in the previous section, so they have no phase transition. Going beyond the independent assumption, it is easy to show by a transfer matrix argument that, if $J_n$ is a time-additive functional of a \emph{time-homogeneous} Markov process with a gapped spectrum, then
\be
U(k,n) \sim (1-r)^n e^{n\lambda_0(k)+\Delta \epsilon^n},
\ee
where $\Delta$ is a constant depending on the initial condition and, crucially, $\epsilon<1$ so we again have $c=0$. Processes that are non-homogeneous in time, however, can lead to generating functions having subleading terms in $n$. In the following we consider three examples in this class in which the current increments, although no longer identically distributed, are still independent. For these weakly time-dependent random walks the currents correspond mathematically to sums of independent but non-identically distributed random variables.

\subsection{Gaussian random walk with varying variance}

We first consider a random walker which, at the $i$th step after resetting, takes a jump drawn from a Gaussian distribution with zero mean and time-dependent variance $A[1-B/(i+d)]$, where $A>0$, and $B$ and $d$ are constants such that $B\leq 1+d$ to ensure positive variance for all steps. The variance goes to the constant $A$ as $i\ra\infty$ for any finite $d$, but taking $d>0$ allows us to explore larger values of $B$.  

For this model, the generating function of the total displacement (viz., current) for $n$ steps without reset is simply  given by
\be
U(k,n)=(1-r)^n\prod_{i=1}^n \exp\left[\frac{A}{2}\left(1-\frac{B}{i+d}\right) k^2 \right]. 
\label{e:UGaussianSD}
\ee
Setting $A=2$, without loss of generality, we can rewrite this as
\be
U(k,n)=(1-r)^n\exp\left[ n k^2  - B k^2 (H_{n+d}-H_d) \right],
\ee
where
\be
H_n=\sum_{k=1}^n \frac{1}{k}
\ee
is the $n$th harmonic number. Using the known asymptotic expression $H_n \approx \ln n + \gamma$, where $\gamma=0.57721\ldots$ is the Euler-Mascheroni constant, yields the scaling
\be
U(k,n) \sim \frac{(1-r)^n e^{nk^2}}{n^{Bk^2}},
\ee
which has the form \eqref{e:Unlongtime} with $\lambda_0(k)=k^2$ and $c(k)=Bk^2$.  We immediately see that a phase transition is only possible for $B>0$ (variance increasing with $i$); in this case there is a competition between the increasing probability of seeing large displacements in longer no-reset segments and the decreasing probability of having fewer resets in the first place.

The $z$-transform of~\eqref{e:UGaussianSD} with $A=2$ is
\be
\tU(k,z)= \sum_{n=1}^\infty (1-r)^n\exp\left[ n k^2  - B k^2 (H_{n+d}-H_d) \right] z^{-n}
\label{e:UtGaussianSD}
\ee
for which no analytic expression can be found for $B>0$. The sum, however, can be used to solve~\eqref{e:zstar} numerically for $z^*(k)$ and to look for phase transitions where it becomes equal to $z_c(k)=(1-r)e^{k^2}$.   Since the SCGF is obviously an even function of $k$, we concentrate here on $k \geq 0$. 

From~\eqref{e:kc} the value of $k_c$ at which a phase transition occurs must satisfy
\be
\sum_{n=1}^\infty \exp\left[ - B k_c^2 (H_{n+d}-H_d) \right] = \frac{(1-r)e^{k_c^2}}{r} - 1. 
\label{e:GaussianSD_kc}
\ee
For $B>0$ and $0<r<1$, both sides of this equation depend monotonically on $k_c^2$; the left-hand side diverges as $k_c \ra 1/\sqrt{B}$ and approaches zero as $k_c \ra \infty$ whereas the right-hand side is finite for $k_c=0$ and diverges as $k_c \ra \infty$.  Hence, for the whole parameter range, there is a single (positive) value of $k_c$ satisfying the equation and marking a phase transition.

\begin{figure}[t]
\centering
\includegraphics{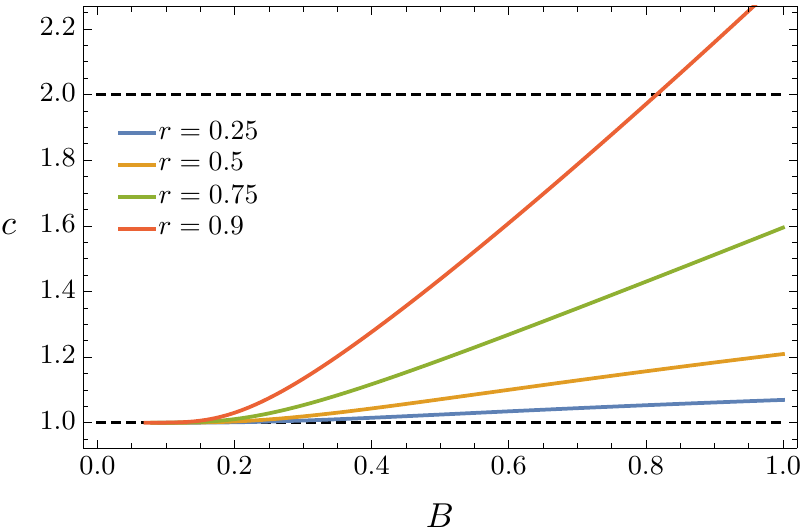}
\caption{(Color online) Exponent $c=Bk_c^2$ at the phase transition point for the Gaussian random walk with varying variance. Parameters: $d=0$ and $r$ values as shown in the legend. As $B\ra 0$, $c\ra 1$. Exponents greater than two indicate first-order dynamical phase transitions.}
\label{figvarpow}
\end{figure}

\begin{figure}[t]
\centering
\includegraphics{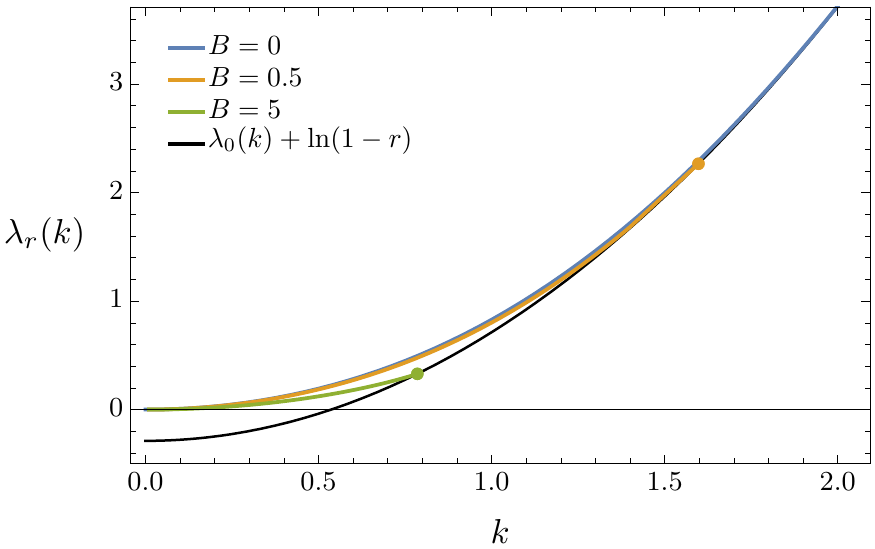}
\caption{(Color online) SCGF $\lambda_r(k)$ for the varying-variance random walk with $d=10$ and $r=0.25$. Coloured lines show~\eqref{eqscgfres1} for different values of $B$ while their intersections with the black line~\eqref{eqscgfnr1} indicate dynamical phase transitions: continuous for $B=0.5$ (yellow circle) and discontinuous for $B=5$ (green circle). The case $B=0$ corresponds to the lazy random walk.}
\label{figvarscgf}
\end{figure}

For $d=0$, the plot of the exponent $c=B k_c^2$ shown in Fig.~\ref{figvarpow} indicates that this phase transition only becomes first-order for large $r$ and large $B$ \footnote{The numerical results in this and subsequent figures were obtained using Maple which appears to implicitly use a zeta-function representation of the left-hand side of~(\ref{e:GaussianSD_kc}) and thus finds a solution even when convergence is slow. We have checked consistency with Mathematica results for parameters that lead to faster convergence.}. A similar analysis holds for $d>0$ and is supported by direct calculations of $\lambda_r(k)$ from~\eqref{eqscgfres1} as shown in Fig.~\ref{figvarscgf}.  Here the analytical curve for $B=0$ (corresponding to a lazy random walk) shows no phase transition, as expected.  The numerical results for $B=0.5$ show that the two solutions (\ref{eqscgfres1}) and (\ref{eqscgfnr1}) meet at $k_c$ with equal derivatives (but different second derivatives), marking a continuous transition between fluctuations that typically involve resets and fluctuations that do not. For $B=5$, (\ref{eqscgfres1}) and (\ref{eqscgfnr1}) meet at a lower $k_c$ with different derivatives, creating a cusp in $\lambda_r(k)$ which marks a discontinuous transition between the reset and no-reset fluctuation regimes.

\begin{figure}[t]
\centering
\includegraphics{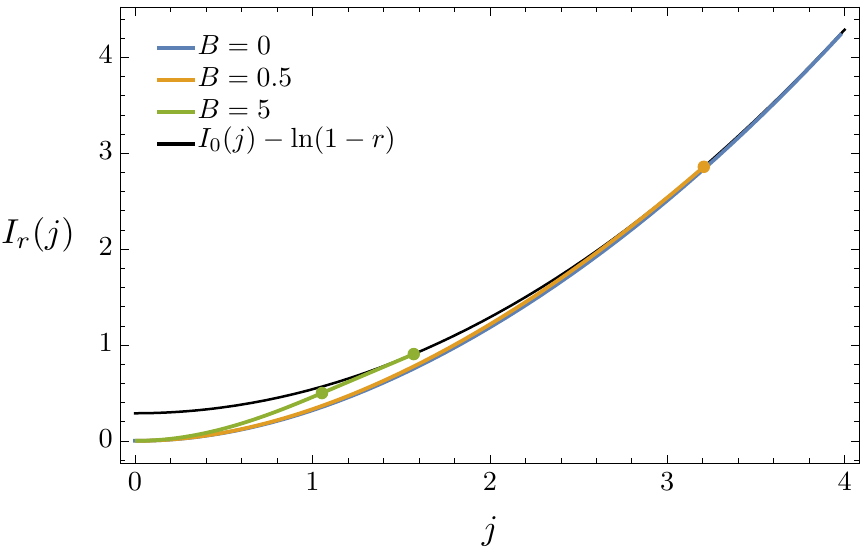}
\caption{Rate function $I_r(j)$ of the varying-variance random walk with $d=10$ and $r=0.25$. The rate function for $B=5$ has a linear section between the two green circles corresponding to the left- and right-derivative, respectively, of $\lambda_r(k)$ at $k_c$.}
\label{figrf}
\end{figure}

The likelihood of each regime is determined from the rate function $I_r(j)$, shown in Fig.~\ref{figrf}, which is obtained by numerically computing the LF transform of the SCGF. For $B=0.5$ the rate function shows the same discontinuity in the second derivative as for the SCGF, whereas for $B=5$ the non-differentiable point of the SCGF transforms into a straight line connecting the reset and no-reset branches \footnote{The rate function is expected to be convex here, since there are no long-range temporal correlations. Therefore, the LF transform should give the correct rate function even if the SCGF is not everywhere differentiable.}. This line is interpreted physically as a mixed regime (``phase separated in time'') where typical trajectories switch between periods with frequent resets and periods with no resets \footnote{This is analogous to the ``Maxwell construction'' in equilibrium statistical mechanics and relies again on the absence of long-range temporal correlations.}. In the case where $r\ra 1$ and $B=1$, it can be checked that the straight line extends to $j=0$ and the rate function approaches
\be
I_r(j)= \begin{cases}
{k_c |j| } & \text{for $|j|\leq2k_c$} \\
{\frac{j^2}{4}-\ln(1-r) } & \text{for $|j|>2k_c$},
\end{cases} 
\label{e:approxLegendre}
\ee
where $k_c=\sqrt{-\ln(1-r)}$. Thus in this case there is a mixed regime of periods with no current flow and periods with non-zero current.  As $j$ increases the fraction of the trajectory occupied by the latter periods increases, yielding exponential fluctuations up to a critical current $j_c=2k_c$ beyond which the fluctuations become Gaussian and involve no reset.

\subsection{Gaussian random walk with decaying mean}

As a variant of the previous model, we now consider step lengths with constant variance but mean $B/i$ for the $i$th step after reset. In this case, the generating function for $n$ steps without reset is
\be
U(k,n)=(1-r)^n\prod_{i=1}^n \exp\left(\frac{A}{2} k^2 +\frac{B}{i} k \right).
\label{e:UGaussianM}
\ee
We concentrate here on $B>0$ corresponding to a positive initial bias decreasing with the step number as $1/i$ \footnote{We could also include a shift $d$, as in the previous example, but since this variant already allows arbitrarily large $B$, we do not pursue that complication here.}.  Setting $A=2$, as before, we then have  
\be
U(k,n)=(1-r)^n e^{nk^2 + B k H_n}
\sim \frac{ (1-r)^n e^{nk^2}}{n^{-Bk}}
\ee
which has the form \eqref{e:Unlongtime} with $\lambda_0(k)=k^2$ again but now $c(k)=-Bk$. Notice here that the exponent in the denominator can be either positive or negative depending on the sign of $k$.  Hence we see that there is no phase transition for positive $k$ (i.e., positive current fluctuations). For negative $k$ there is a phase transition at $k_c$ satisfying
\be
\sum_{n=1}^\infty e^{B k_c H_{n}} = \frac{(1-r)e^{k_c^2}}{r} - 1, 
\label{e:GaussianM_kc}
\ee
as obtained from~\eqref{e:kc} with the $z$-transform of ~\eqref{e:UGaussianM}.  Again, it is easy to argue that this equation has a single solution for all $B>0$ and $0<r<1$.

\begin{figure}[t]
\centering
\includegraphics{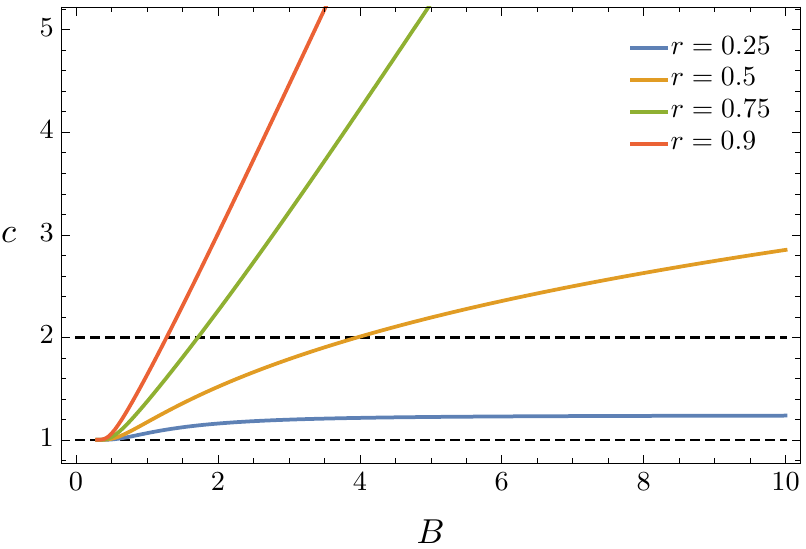}
\caption{(Color online) Exponent $c=-Bk_c$ at the phase transition point for the Gaussian random walk with decaying mean and $r$ values as shown in the legend. As $B\ra0$, $c\ra 1$. Exponents greater than two indicate  first-order dynamical phase transitions.}
\label{figmeanpow}
\end{figure}

\begin{figure}[t]
\centering
\includegraphics{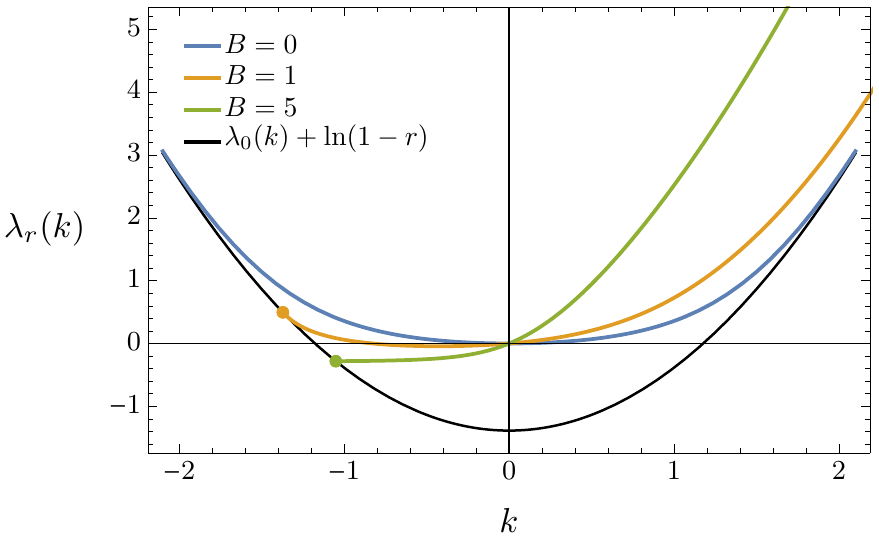}
\caption{(Color online) SCGF $\lambda_r(k)$ of the Gaussian random walk with decaying mean and $r=0.75$. A continuous dynamical phase transition is seen for $B=1$, while a first-order transition is seen for $B=5$.}
\label{figmeanscgf}
\end{figure}

In Fig.~\ref{figmeanpow} we plot numerical results for $c=-B k_c$ against $B$ for different values of $r$. When $B \ra 0$, we get $k_c \ra -\infty$ and,  as in the previous model, $c \ra 1$. The present model also allows us to easily explore the limiting behaviour when $B \ra \infty$ which turns out to depend qualitatively on $r$. For $r > 1/2$, $k_c \ra -\sqrt{\ln[r/(1-r)]}$ and $c$ has an oblique asymptote. In contrast for $r \leq 1/2$, $k_c \ra 0$ and $c$ approaches a constant; if this constant is less than two, the phase transition remains continuous however large $B$ is. We have checked these predictions by numerically calculating the SCGF via solution of~\eqref{e:zstar}. In Fig.~\ref{figmeanscgf} we plot the results for $r=0.75$ showing first-order and continuous phase transitions in the current fluctuations for different values of $B$, similarly to the previous model. The associated rate function is also similar to that of Fig.~\ref{figrf} and is not shown for this reason.

One important difference to the previous model is that, although the SCGF is still even in $k$ without reset (because the step mean decays to zero in the long-time limit), the addition of reset breaks this symmetry, bringing a non-zero (positive) mean current. The behaviour in the large-$B$ limit is also interesting: from the discussion of $k_c$ in the previous paragraph we find that, for $r>1/2$, $\lambda_r(k_c) \to \ln r$ whereas, for $r \leq 1/2$, $\lambda_r(k_c) \to \ln (1-r)$.  In the former case, $z_c$ approaches the convergence boundary point $r$ of $\tV(k,z)$ and the effect of this pole can already be seen in the almost flat part of the $B=5$ line in Fig.~\ref{figmeanscgf}.  In fact, the structure of the right-hand side of~\eqref{e:kc} and the form of $z_c$ suggest that the distinction between $r<1/2$ and $r>1/2$ should be rather generic.

\subsection{Discrete random walk with decaying mean}

For our last example, we briefly consider a random walk on a one-dimensional lattice with transition probabilities that are weakly asymmetric in time in the sense that, at the $i$th time step after reset, the random walker moves one lattice unit right with probability $[1+B/(i+d)]/2$ and one lattice unit left with probability $[1-B/(i+d)]/2$ where $0\leq B \leq 1+d$. In the context of opinion dynamics, this can be thought of as a discrete-choice model where an agent's bias decays with time until reset by some particular event. 

For small $k$, this model behaves similarly to the Gaussian random walk with varying mean (the steps have mean $B/(i+d)$ and unit variance), but is notably simpler to analyse analytically. The current generating function for $n$ steps without reset is here
\be
U(k,n)= [(1-r)\cosh k]^n \frac{(B\tanh k +d+1)_n}{(d+1)_n},
\label{e:UD}
\ee
where $(x)_n$ is the Pochhammer symbol defined in terms of the Gamma function as $\Gamma(x+n)/\Gamma(x)$.  This yields the asymptotic behaviour
\be
U(k,n) \sim \frac{1}{n^{-B\tanh k}} [(1-r)\cosh k]^n
\ee
corresponding to~\eqref{e:Unlongtime} with $\lambda_0(k)=\ln(\cosh k)$ and $c(k)=-B\tanh k$.  For small $k$, the exponent $c(k)$ is unsurprisingly close to that of the Gaussian decaying-mean model, but it differs for large $k$ since here $c(k) \ra B$ as $k \ra - \infty$.  This means that for $B \leq 1$ there can be no phase transition at any finite value of $k$ regardless of the value of $r$. In particular, there is no phase transition for $d=0$ where $B \leq 1$ by construction.

This result can be verified explicitly for $d=0$ since in this case $\tU(k,z)$ takes the simple analytic form
\be
\tU(k,z) = -1 + [1 - (1 - r) z^{-1} \cosh k]^{-B \tanh k - 1}.
\ee
One sees directly that $z_c(k)=(1-r)\cosh k$ and $\tU(k,z_c)$ diverges, which means that  \eqref{e:kc} cannot be satisfied for any finite $k_c$.  For $d>0$, there also exists an analytic expression for $\tU(k,z)$ in terms of the hypergeometric function, which predicts a phase transition if $-B \tanh k_c > 1$ independently of $d$, in agreement with the earlier analysis. 

\section{Conclusions}

We have shown that dynamical phase transitions can arise in the fluctuations of time-integrated observables of reset processes, following a mechanism analogous to how phase transitions arise in the Poland-Scheraga (PS) model of DNA denaturation. In such processes, subexponential terms in the generating function of the observable, which play no role in the long-time limit without reset, are ``amplified'' by the presence of reset, leading to continuous and discontinuous transitions between reset and no-reset fluctuation regimes.

Following the random walk examples presented here, we expect similar dynamical phase transitions to arise in many other settings, including more general ``compound processes'' that switch at random times between two or more independent processes. In this context, it would be interesting to investigate continuous-time models (which do not have such a direct mapping to the PS model), systems with time-dependent reset or switching events, as in~\cite{pal2016, nagar2016}, and non-Markovian dynamics having transition rates that depend on the whole history of the process (see, e.g,~\cite{harris2009,harris2015}). For illustrative purposes, we have restricted ourselves to models with zero mean current in the absence of reset, but the analysis can also be extended to driven non-equilibrium systems where we anticipate that reset-induced dynamical phase transitions will break the Gallavotti-Cohen symmetry~\cite{lebowitz1999} for the current of the original dynamics.

Finally, it is possible to investigate the joint statistics of reset events and currents via the joint generating function of the current $J_n$ and the number $R_n$ of resets after $n$ time steps. Due to the structure of our problem, this simply amounts to replacing $r$ by $r e^l$ on the right-hand side of~\eqref{e:zstar}, with $l$ as the conjugate parameter associated to $R_n$. The solution $z^*$ then becomes a function of both $k$ and $l$, yielding a joint generating function and, by Legendre-Fenchel transform, a joint rate function. The value of $R_n/n$ minimizing this rate function for a given $J_n/n$ corresponds to the optimal way to realise that current fluctuation and thus illuminates the physical structure underlying any dynamical phase transitions.

\begin{acknowledgments}
R.J.H.\ thanks Bernard Derrida for introducing her to the Poland-Scheraga model and also gratefully acknowledges NITheP for a funded research visit. H.T.\ is supported by the  National Research Foundation of South Africa (Grants no.\ 90322 and 96199) and Stellenbosch University (Project Funding for New Appointee).
\end{acknowledgments}

\bibliography{masterbibmin}

\end{document}